\documentclass[journal,onecolumn,12pt]{IEEEtran}
\usepackage{cite}   
\usepackage{amsmath,amssymb}
\usepackage{enumerate}
\usepackage{array}
\newtheorem{theorem}{Theorem}[section]
\newtheorem{corollary}[theorem]{Corollary}\newtheorem{lemma}[theorem]{Lemma}
\begin{document}
\title{On Quadratic Inverses for Quadratic Permutation Polynomials over Integer Rings}

\author{Jonghoon Ryu and Oscar Y. Takeshita  \\
Dept. of Electrical and Computer Engineering \\
2015 Neil Avenue \\
The Ohio State University \\
Columbus, OH 43210 \\ 
\{ryu.38, takeshita.3\}@osu.edu\\
\vspace{2em}
Submitted as a Correspondence to the IEEE Transactions on Information Theory\\
Submitted : April 1, 2005\\
Revised : Nov. 15, 2005}

\maketitle
\begin{abstract}
  An interleaver is a critical component for the channel coding
  performance of turbo codes. 
  Algebraic constructions are of particular interest because they
  admit analytical designs and simple, practical hardware
  implementation.  Sun and Takeshita have recently shown
  that the class of quadratic permutation polynomials over integer
  rings provides excellent performance for turbo codes.  In this
  correspondence, a necessary and sufficient condition is proven for the
  existence of a quadratic inverse polynomial for a quadratic
  permutation polynomial over an integer ring.  Further, a simple
  construction is given for the quadratic inverse.  All but one of the
  quadratic interleavers proposed earlier by Sun and Takeshita are found to admit
  a quadratic inverse, although none were explicitly designed to do
  so.  An explanation is argued for the observation that restriction
  to a quadratic inverse polynomial does not narrow the pool of good
  quadratic interleavers for turbo codes. 
\end{abstract}
\pagebreak
\section{Introduction}
  Interleavers for turbo codes~\cite{Sun, Berrou, Sadjadpour, Takeshita1,Takeshita2, Dolinar, Divsalar, Bravo, Daneshgaran ,Crozier1,Crozier2}
  have been extensively investigated. 
  Recently, Sun and Takeshita\cite{Sun} suggested the use of permutation
  polynomial-based interleavers over integer rings. 
  In particular, quadratic polynomials were emphasized; 
  this quadratic construction is markedly
  different from and superior to the one proposed earlier by Takeshita and Costello~\cite{Takeshita1}.
  The algebraic approach was shown to admit analytical design of an
  interleaver matched to the constituent convolutional codes.
  The resulting performance was shown to be better than S-random that of interleavers~\cite{Dolinar, Divsalar} for
  short to medium block lengths and parallel concatenated turbo codes~\cite{Sun,Takeshita2}. 
  Other interleavers~\cite{Crozier1, Crozier2} better than S-random interleaver
  for parallel concatenated turbo codes have also been investigated but they are not algebraic.  \\
  This correspondence is motivated by  work at the Jet Propulsion Laboratory (JPL) \cite{JPL1, JPL2} for the Mars Laser
  Communication Demonstration (MLCD). 
  The interleaver in \cite{JPL2} is used in a serially concatenated turbo code.
  The work in \cite{JPL2} shows that the quadratic interleavers proposed in \cite{Sun}
  can be efficiently implemented in Field-Programmable Gate Array
  (FPGA) using only additions and comparisons. A turbo decoder needs also a deinterleaver.
  In \cite{JPL1}, the inverse polynomial of a quadratic polynomial is 
  computed by brute force using the fact that permutation polynomials form a group
  under function composition. It is also shown that the inverse
  polynomial of a quadratic permutation polynomial may not be quadratic by a particular counterexample. 
  Therefore two natural questions arise: 
  When does a quadratic permutation polynomial over
  an integer ring have a quadratic inverse polynomial? How do we compute it efficiently? \\ \indent
  In this correspondence, we derive a necessary and sufficient condition for a
  quadratic permutation polynomial over integer rings to admit a
  quadratic inverse. The condition consists of simple arithmetic comparisons.
  In addition, we provide a simple algorithm to compute the inverse
  polynomial. Further, we argue that this restriction does not seem to
  effectively narrow the pool of good quadratic interleavers for turbo
  codes. \\
%
%
  This correspondence is organized as follows. 
  In section II, we briefly review permutation polynomials\cite{Hardy, Rivest, Mullen1, Lidl1, Lidl2, Lidl3} 
  over the integer ring $\mathbb{Z}_N$
  and relevant results. The main result is derived in section III, and
  examples are given in section IV. Finally, conclusions are discussed in section V. 
\section{Permutation Polynomial over Integer Rings}
  In this section, we revisit the relevant facts about permutation
  polynomials and other additional results in number theory to make this correspondence self-contained.\\ 
  Given an integer $N \geq 2$, a polynomial $\overline{H}(x) = h_0 + h_1 x + h_2 x^2 + \cdots +h_k  x^k \pmod{N}$, 
  where $h_0, h_1, \ldots,h_k$ and $k$ are non-negative integers, is said to be a permutation polynomial over
  $\mathbb{Z}_N$ when $\overline{H}(x)$ permutes $\{0, 1, 2, \ldots, N-1 \}$\cite{Rivest,Mullen1, Lidl1, Lidl2, Lidl3}. 
  It is immediate that the constant $h_0$ in $\overline{H}(x)$ only causes a ``cyclic shift'' to the permuted values.
  Therefore we define the polynomial $H(x) = \overline{H}(x)-h_0$ without losing generality in our quest
  for a quadratic inverse polynomial by the following lemma.
\begin{lemma}
  Suppose that the inverse of a permutation polynomial $H(x)$ is $I(x)$.
  Then the inverse permutation polynomial of $\overline{H}(x)$ is $I(x-h_0)$. 
  Conversely, suppose that the inverse of a permutation polynomial $\overline{H}(x)$ is $J(x)$.
  Then the inverse permutation polynomial of $H(x)$ is $J(x+h_0)$.
\end{lemma}
\begin{proof}
  Suppose the inverse of $H(x)$ is $I(x)$. Then $H(I(x)) = x$. 
  Consequently, $\overline{H}(I(x-h_0))=H(I(x-h_0))+h_0 = x-h_0+h_0 = x$.
  The other direction can be proved similarly. 
\end{proof}
Further, it is well known that an inverse permutation polynomial always exists
because permutation polynomials forms a group under function
composition~\cite{JPL1, Lidl1, Lidl2, Lidl3}. 
The condition for a quadratic polynomial to be a permutation polynomial over $\mathbb{Z}_p$, 
where $p$ is any prime, is shown in the following two lemmas. 
\begin{lemma}[\cite{Rivest}]
  Let $p = 2$. A polynomial $H(x) = h_1x + h_2 x^2 \pmod{p}$ is a permutation polynomial over 
  $\mathbb{Z}_{p}$ if and only if $h_1+ h_2$ is odd. 
\end{lemma}
\begin{lemma}[\cite{Lidl1}]
  Let $p \neq 2$.  A polynomial $H(x) = h_1x + h_2 x^2 \pmod{p}$ is a permutation polynomial over 
  $\mathbb{Z}_{p}$   if and only if 
  $h_1 \not\equiv 0 \pmod{p} $ and $h_2 \equiv 0 \pmod{p}$, i.e., there are no quadratic permutation polynomials modulo a prime $p\neq 2$.
\end{lemma}
The following theorem and corollary give the necessary and sufficient conditions for a polynomial to be a permutation polynomial 
over integer ring $\mathbb{Z}_{p^n}$, where $p$ is any prime number and  $n \geq 2$. 
\begin{theorem}[\cite{Hardy, Sun}]
  Let $p$ be a prime number and $n \geq 2$. A polynomial $H(x) = h_1x + h_2 x^2 \pmod{p^n}$ is a permutation polynomial over  
  $\mathbb{Z}_{p^n}$ if and only if 
  $h_1 \not\equiv 0 \pmod{p} $ and $ h_2 \equiv 0 \pmod{p}$.
\end{theorem}
\begin{corollary}[\cite{Rivest}]
  Let $p=2$ and $n \geq 2$. A polynomial $H(x) = h_1 x + h_2 x^2 \pmod{p^n}$ is a permutation polynomial 
  if and only if $h_1$ is odd and $h_2$ is even.
\end{corollary}
Corollary 2.5 can be easily verified from Theorem 2.4. However, since our proofs in the Appendix can be simplified using Corollary 2.5, 
we keep it for its simplicity. 
In this correspondence, let the set of primes be $\mathcal{P} = \{2, 3,  5 , ...   \}$. 
Then an integer $N$ can be factored as $N = \prod\nolimits_{p \in \mathcal{P}}  p^{n_{N,p}}  $, 
where $p$'s are distinct primes, $n_{N,p} \geq 1$ for a finite number of $p$ and $n_{N,p}=0$ otherwise. 
For a quadratic polynomial $H(x)=h_1 x + h_2 x^2 \pmod{N}$, we will abuse the
previous notation by writing $h_2 = \prod\nolimits_{p \in \mathcal{P}}  p^{n_{H,p}} $, i.e., the exponents of the
prime factors of $h_2$ will be written as $n_{H,p}$ instead of the more
cumbersome $n_{{h_2},p}$ because we will only be interested in the
factorization of the second degree coefficient. \\
For a general $N$, the necessary and sufficient condition for a polynomial to be a permutation polynomial is given in the following theorem. 
\begin{theorem}[\cite{Sun}]
  For any $N = \prod\nolimits_{p \in \mathcal{P}}  p^{n_{N,p}}  $, $H(x)$ is a permutation
  polynomial modulo $N$ if and only if $H(x)$ is also a permutation polynomial modulo $p^{n_{N,p}} $, 
  $\forall p$ such that  $n_{N,p} \geq 1$. 
\end{theorem}
Using this theorem, verifying whether a polynomial is a permutation polynomial modulo $N$
reduces to verifying the polynomial modulo each $p^{n_{N,p}} $ factor of $N$.
\begin{corollary}
  Let $N = \prod\nolimits_{p \in \mathcal{P}}  p^{n_{N,p}} $ and denote $y$ divides $z$ by $y|z$. 
  The necessary and sufficient condition for a quadratic polynomial $H(x) = h_1 x + h_2 x^2 \pmod{N}$ 
  to be a permutation polynomial can be divided into two cases. 
  \begin{enumerate}
    \item $2|N$ and $4\nmid N$ (i.e., $n_{N,2}=1$)\\
    $h_1+h_2$ is odd, $\gcd(h_1,\frac{N}{2}) = 1$ and $h_2 = \prod\nolimits_{p \in \mathcal{P}}  p^{n_{H,p}}, n_{H,p} \geq 1$, $\forall p$ 
    such that $p \neq 2$ and $n_{N,p} \geq 1$.
    \item Either $2 \nmid N$ or $4 |N$ (i.e., $n_{N,2}\not = 1$)\\
    $\gcd(h_1, N)=1$ and $   h_2 = \prod\nolimits_{p \in \mathcal{P}}  p^{n_{H,p}}, n_{H,p} \geq 1  $, $\forall p$
    such that  $n_{N,p} \geq 1$.
  \end{enumerate}
\end{corollary}
\begin{proof}
  This is a direct consequence of Lemmas 2.2, 2.3, Theorems 2.4, 2.6 and Corollary 2.5. 
\end{proof}
The following theorem and lemma are also necessary for deriving the main theorem (Theorem 3.6) of this correspondence. 
\begin{theorem}[\cite{Hardy}]
  Let $a$, $b$ and $N$ be integers. The linear congruence $au \equiv b \pmod{N}$ has at least one solution if and only if $d | b$, where $d = \gcd(a,N)$.
  If $d | b$, then it has  $d$ mutually incongruent solutions. Let $u_0$ be one solution, then the set of 
  the solutions is  $$u_0, u_0+\frac{N}{d}, u_0+\frac{2N}{d}, \dots, u_0+\frac{(d-1)N}{d},$$ where $u_0$ is the unique solution of 
  $\frac{a}{d}u \equiv \frac{b}{d} \pmod{\frac{N}{d}}$. 
\end{theorem}
\begin{lemma}[\cite{Hardy}]
  Let $M$ be an integer. Suppose that $M | N$ and that $v \equiv w \pmod{N}$. Then $v \equiv w \pmod{M}$.
\end{lemma}
\section{Quadratic Inverse Polynomial}
  In this section, we derive the necessary and sufficient condition for a quadratic polynomial to admit at least one
  quadratic inverse in Theorem 3.6. We also explicitly find the quadratic inverse in Algorithm 1.
  If $N=2$, the inverse polynomial of a quadratic permutation polynomial can be easily constructed. If $N \neq 2$ is a prime number,
  by Lemma 2.3, there are no quadratic permutation polynomials. If $N$ is a composite number,
  it has been shown that the inverse polynomial may not be quadratic by a particular counterexample\cite{JPL1}. 
  However, in the following lemma, it is shown that for any quadratic permutation polynomial, there exists at least one quadratic polynomial that
  inverts it at three points $x=0, 1, 2$. The reason why we look at this partially inverting polynomial is
  because it becomes the basis for the true quadratic inverse polynomial if it exists.  
  \begin{lemma}
    Let $N$ be a composite number and let $F(x) = f_1 x + f_2 x^2 \pmod{N}$ be a quadratic permutation polynomial. 
    Then there exists at least one quadratic polynomial 
    $G(x) = g_1x + g_2 x^2 \pmod{N} $ that inverts  $F(x)$
    at these three points: $x = 0, 1, 2$.
    If $N$ is odd, there
    is exactly one quadratic polynomial $G(x) = g_1x + g_2 x^2 \pmod{N} $ and the coefficients of the polynomial can be obtained by solving the
    linear congruences.
    \begin{eqnarray} 
      \label{eq:06a}   g_2 (f_1+f_2)(f_1 + 2 f_2)(f_1+3 f_2)  \equiv - f_2 \pmod{N}.  \\
      \label{eq:06b}    g_1 (f_1 + f_2) + g_2 (f_1 + f_2)^2 \equiv 1 \pmod{N}. 
    \end{eqnarray}
    If $N$ is even, there
    are exactly two quadratic polynomials  $G_1(x) = g_{1,1}x + g_{1,2} x^2 \pmod{N}$, 
    $G_2(x) = g_{2,1}x + g_{2,2} x^2 \pmod{N}$  and the coefficients of the polynomial
    $G_1(x) = g_{1,1}x + g_{1,2} x^2 \pmod{N}$ can be obtained by solving the linear congruences.
    \begin{eqnarray} 
      \label{eq:06c} g_{1,2}(f_1+f_2)(f_1 + 2 f_2)(f_1+3 f_2)  \equiv -f_2  \pmod{\frac{N}{2}}.  \\ 
      \label{eq:06d}      g_{1,1} (f_1 + f_2) + g_{1,2} (f_1 + f_2)^2 \equiv 1 \pmod{N}. 
    \end{eqnarray}
    After computing $(g_{1,1},g_{1,2})$, $(g_{2,1},g_{2,2})$ can be obtained by 
    $ g_{2,1} \equiv g_{1,1} + \frac{N}{2}  \pmod{N} $ and  $ g_{2,2} \equiv g_{1,2} + \frac{N}{2}  \pmod{N} $.
  \end{lemma}
  \begin{proof}
    See Appendix A. 
  \end{proof}
  Each of the above four linear congruences (\ref{eq:06a}), (\ref{eq:06b}), (\ref{eq:06c}) and (\ref{eq:06d}) are
  guaranteed to have exactly one solution by Lemma 3.1 and Theorem 2.8. It
  is well known that linear congruences can be efficiently solved by using the
  extended Euclidean algorithm~\cite{Gathen}. For example, in congruence (\ref{eq:06b}), the
  unknown value to be solved is $g_1$; $f_1$ and $f_2$ are given and $g_2$
  can be calculated from (\ref{eq:06a}). The congruence (\ref{eq:06b}) can be rewritten as
  \begin{eqnarray}
     g_1\equiv (f_1+f_2)^*\cdot(1-g_2(f_1+f_2)^2) \pmod{N},
  \end{eqnarray}
  where $(f_1+f_2)^*$ means the arithmetic inverse of $(f_1+f_2)
  \pmod{N}$.  By an arithmetic inverse of $s$ modulo
  $N$, we mean a number $s^*$ such that $ss^* \equiv 1 \pmod{N}$.
  The Algorithm 2 provided in Table~\ref{tab:02} can be used
  to calculate such an inverse. \\
  In the following lemma, we show that the polynomials $G(x)$, $G_1(x)$ and $G_2(x)$ obtained by 
  solving the congruences  (\ref{eq:06a}), (\ref{eq:06b}), (\ref{eq:06c}) and (\ref{eq:06d}) are permutation
  polynomials. 
  \begin{lemma}
    The polynomials $G(x)$, $G_1(x)$ and $G_2(x)$ obtained in Lemma 3.1 are permutation polynomials. 
  \end{lemma}
  \begin{proof}
    See Appendix B. 
  \end{proof}
  From Lemmas 3.1 and 3.2, there exists at least one quadratic permutation polynomial $G(x)$ that inverts any quadratic
  permutation polynomial $F(x)$ at three points $x=0,1,2$. However, it does not necessarily mean that $G(x)$ is an inverse polynomial of $F(x)$. \\
  In the following lemma, we show that some exponents $n_{G,p}$'s of the $g_2$ which was obtained in Lemma 3.1  are determined by the exponents
  $n_{F,p}$'s. 
  \begin{lemma}
    Let $N = \prod\nolimits_{p \in \mathcal{P}}  p^{n_{N,p}} $, $F(x) = f_1x + f_2 x^2 \pmod{N} $ 
    be a quadratic permutation polynomial and $G(x) = g_1x + g_2 x^2 \pmod{N} $ be a quadratic permutation polynomial in Lemmas 3.1 and 3.2.   
    Then, $f_2 = \prod\nolimits_{p \in \mathcal{P}}  p^{n_{F,p}} $ and $g_2 = \prod\nolimits_{p \in \mathcal{P}}  p^{n_{G,p}} $
    satisfy Corollary 2.7. Furthermore, the following holds. 
    \begin{enumerate}[{case} a:]
       \item $2 \nmid N $ (i.e., $n_{N,2}=0$)  \\
         If $N$ contains $p$ as a factor (i.e., $n_{N,p}\geq 1$) then
         \[
         \left\{
            \begin{array}{ccc}
              n_{G,p} = n_{F,p} & \mbox{if} & 1 \leq n_{F,p} < n_{N,p} \\
              n_{G,p} \geq n_{N,p} & \mbox{if}  & n_{F,p} \geq n_{N,p}
            \end{array}
         \right.
         \]
       \item $ 2 | N$ and $ 4 \nmid N $ (i.e., $n_{N,2}=1$) \\
         $N$ contains $p =2 $ as a factor but we do not need to consider how
         $n_{G,2}$ is determined by $n_{F,2}$. The reason for this is explained in the proof of Theorem 3.6. \\
         If $p \neq 2$ and $N$ contains $p$ as a factor (i.e., $n_{N,p}\geq 1$)  then
         \[
         \left\{
            \begin{array}{ccc}
               n_{G,p} = n_{F,p} & \mbox{if} & 1 \leq  n_{F,p} < n_{N,p} \\
               n_{G,p} \geq n_{N,p} & \mbox{if}  & n_{F,p} \geq n_{N,p}
            \end{array}
         \right.
         \]
       \item $ 4 | N $ (i.e., $n_{N,2} \geq 2$)\\
          $N$ contains $2^2$ as a factor (i.e., $n_{N,2}\geq 2$). \\
         If $p = 2$, 
          \[
          \left\{
             \begin{array}{ccc}
                n_{G,2} = n_{F,2} & \mbox{if} & 1 \leq n_{F,2} < n_{N,2}-1 \\
                n_{G,2} \geq n_{N,2}-1 & \mbox{if}  & n_{F,2} \geq n_{N,2}-1
             \end{array}
          \right.
          \]
         If $p \neq 2$ and $N$ contains $p$  as a factor (i.e., $n_{N,p}\geq 1$) then
          \[
          \left\{
             \begin{array}{ccc}
                n_{G,p} = n_{F,p} & \mbox{if} & 1 \leq n_{F,p} < n_{N,p} \\
                n_{G,p} \geq n_{N,p} & \mbox{if}  & n_{F,p} \geq n_{N,p}
             \end{array}
          \right.
         \]
     \end{enumerate}
   \end{lemma}
  \begin{proof}
    See Appendix C. 
  \end{proof}
Before proceeding further, we need the following lemma. 
\begin{lemma}
  Let $T(x) = t_1 x + t_2 x^2 + t_3 x ^3 + t_4 x^4 \pmod{N}$ and $T(0) \equiv T(1) \equiv T(2) \equiv 0 \pmod{N}$.
  Then $T(x)\equiv 0 \pmod{N} $, $\forall x \in [0,N-1]$ if and only if
  $24 t_4 \equiv 0 \pmod{N} $ and $ 6 t_3 + 36 t_4  \equiv 0 \pmod{N}  $.
\end{lemma}
\begin{proof}
  See Appendix D. 
\end{proof}
Combining Lemmas 3.1 and 3.4 gives the following theorem.
\begin{theorem}
  Let   $F(x)$ be a quadratic permutation polynomial and let $G(x)$ be a quadratic polynomial in Lemma 3.1. 
  Then  $G(x)$ is a quadratic inverse polynomial of $F(x)$ if and only if $12f_2g_2 \equiv 0 \pmod{N}$.
\end{theorem}
\begin{proof}
  See Appendix E. 
\end{proof}
We now state our main theorem. It states that the necessary and sufficient condition for the existence of a quadratic inverse for a quadratic permutation polynomial 
$F(x)$ can be simply checked by inequalities involving the exponents for the prime factors of $N$ and the second degree coefficient
of $F(x)$.  
\begin{theorem}[main Theorem]
  Let $N = \prod\nolimits_{p \in \mathcal{P}}  p^{n_{N,p}}$, 
  $F(x)$ be a quadratic permutation polynomial and $f_2 =  \prod\nolimits_{p \in \mathcal{P}}  p^{n_{F,p}} $ be the second degree coefficient of $F(x)$.
  Then $F(x)$ has at least one quadratic inverse polynomial if and only if 
  \[  n_{F,2} \geq \left\{ \begin{array}
              {r@{\quad \mbox{if}\quad}l}
              \max\left(\left\lceil \frac{n_{N,2}-2}{2} \right\rceil,1\right)    &    n_{N,2} > 1  \\
                                                0    &    n_{N,2} = 0,1
              \end{array} \right.   \; , \]  
  \[  n_{F,3} \geq \left\{ \begin{array}
              {r@{\quad\mbox{if}\quad}l}
              \max\left(\left\lceil \frac{n_{N,3}-1}{2} \right\rceil,1\right)    &    n_{N,3} > 0  \\
                                                    0     &    n_{N,3} = 0
              \end{array} \right.   \; ,  \]
  $$  n_{F,p} \geq  \left\lceil \frac{n_{N,p}}{2} \right\rceil \;\; \mbox{if} \;\; p \neq 2, 3.    $$
\end{theorem}
\begin{proof}
  See Appendix F. 
\end{proof}
An interesting question of practical significance is if an interleaver
can be its own inverse~\cite{Takeshita1} because the same hardware can
be used for both interleaving and deinterleaving. It is shown 
in~\cite{Takeshita1} that this type of restriction did not affect turbo
decoding performance using interleavers therein proposed.
Unfortunately, we were not able to identify good self-inverting quadratic permutation polynomials for turbo codes.
%
%
%
%
%
\subsection{Algorithms for Finding the Quadratic Inverse Polynomials}
Algorithm 1 is provided in Table~\ref{tab:01}. It finds the quadratic
inverse polynomial for a given quadratic permutation polynomial $F(x)=f_1x+f_2x^2\pmod N$. In 
Table~\ref{tab:02}, Algorithm 2 is provided to calculate the arithmetic
inverse of $s \pmod {M}$, which is required in Algorithm 1. 
\begin{table}[hbt]
\caption{Algorithm 1}
\begin{center}
\begin{small}
\begin{tabular}{|l|}
\hline
\quad\quad\quad\quad\quad\quad\quad An algorithm for finding the quadratic inverse permutation polynomial(s) \\
\quad\quad\quad\quad\quad\quad\quad\quad     for a quadratic permutation polynomial $F(x) = f_1 x+ f_2 x^2 \pmod{N}$\\ 
\hline
1. Factor $N$ and $f_2$ as products of prime powers and  find the respective exponents of each prime factor. \\
\quad\quad i.e., find $n_{N,p}$'s and $n_{F,p}$'s for  $N = \prod\nolimits_{p \in \mathcal{P}}  p^{n_{N,p}}$, $f_2 = \prod\nolimits_{p \in \mathcal{P}}  p^{n_{F,p}}$,\\
2. Using the $n_{N,p}$'s and $n_{F,p}$'s obtained above, determine if they satisfy the inequalities in Theorem 3.6.  \\
\quad\quad                       if yes, check if $N$ is an odd number, \\
\quad\quad\quad\quad                 if $N$ is an odd number, \\
\quad\quad\quad\quad\quad\quad           There is exactly one quadratic inverse for $F(x)$. \\
\quad\quad\quad\quad\quad\quad           Let the inverse polynomial be $G(x)=g_1x+g_2x^2 \pmod{N}$. \\
\quad\quad\quad\quad\quad\quad           $g_2 \equiv \{[(f_1+f_2)(f_1+2f_2)(f_1+3f_2)]^* \cdot (-f_2)\} \pmod{N}  $, where $(\; )^*$ is given in Algorithm 2.\\
\quad\quad\quad\quad\quad\quad           $g_1 \equiv [(f_1+f_2)^* \cdot  (1-g_2(f_1+f_2)^2)]\pmod{N} $.\\
\quad\quad\quad\quad\quad\quad           Return $G(x)$ and the algorithm ends. \\
\quad\quad\quad\quad                 else $N$ is an even number, \\
\quad\quad\quad\quad\quad\quad           There are exactly two quadratic inverses for $F(x)$. \\
\quad\quad\quad\quad\quad\quad           Let the two inverse  polynomials be $G_1(x)=g_{1,1}x+g_{1,2}x^2 \pmod{N}$ and\\ 
\quad\quad\quad\quad\quad\quad        $G_2(x)=g_{2,1}x+g_{2,2}x^2\pmod{N} $, respectively.\\ 
\quad\quad\quad\quad\quad\quad           $g_{1,2} \equiv \{[(f_1+f_2)(f_1+2f_2)(f_1+3f_2)]^* \cdot (-f_2)\} \pmod{\frac{N}{2}}$\\
\quad\quad\quad\quad\quad\quad           $g_{1,1} \equiv [(f_1+f_2)^* \cdot (1-g_{1,2}(f_1+f_2)^2)]\pmod{N} $.\\
\quad\quad\quad\quad\quad\quad           ($g_{2,1}$, $g_{2,2}$) is obtained by $g_{2,1} \equiv g_{1,1} + \frac{N}{2} \pmod{N}$, $g_{2,2} \equiv g_{1,2} + \frac{N}{2} \pmod{N}$.\\ 
\quad\quad\quad\quad\quad\quad          Return $G_1(x)$ and $G_2(x)$ and the algorithm ends. \\
\quad\quad\quad\quad                 end \\
\quad\quad                       else \\   
\quad\quad\quad\quad                  There exists no quadratic inverse polynomial for $F(x)$.\\
\quad\quad\quad\quad                  The algorithm returns no polynomial and ends. \\
\quad\quad                       end \\
\hline
\end{tabular}
\end{small}
\end{center}
\label{tab:01}
\end{table}
\begin{table}[hbt]
\caption{Algorithm 2}
\begin{center}
\begin{small}
\begin{tabular}{|l|}
\hline
 An algorithm for finding the arithmetic inverse $s^*$
 for $s \pmod{M}$\\
\hline
\quad\quad        $s^* = 1; $ \\ 
\quad\quad        $r = 0; $ \\
\quad\quad        while $(M \neq 0)  $  \\
\quad\quad\quad\quad     $   c \equiv s \pmod{M};$\\
\quad\quad\quad\quad     $   quot = \lfloor \frac{s}{M} \rfloor   ;$\\ 
\quad\quad\quad\quad     $   s = M;$\\
\quad\quad\quad\quad     $ M = c;$\\
\quad\quad\quad\quad     $ r^\prime = s^* - quot * r;$\\ 
\quad\quad\quad\quad     $s^* = r; $\\
\quad\quad\quad\quad     $r =r^\prime;$\\
\quad\quad        end\\
\quad\quad           Return $s^*$\\
 \hline
\end{tabular}
\end{small}
\end{center}
\label{tab:02}
\end{table}
\section{Examples}
We present three examples to illustrate the necessary and sufficient
conditions of Theorem 3.6.
The first example considers interleavers that are now being
investigated in \cite{JPL2}. The second example is a generalization of an
example in \cite{JPL1}. The third example shows that the verification
procedure simplifies when $N$ is a power of 2, 
as it was chosen in~\cite{Sun}, for a fair comparison with~\cite{Takeshita1}.
Remarkably, all good quadratic interleavers 
found in \cite{Sun} except one admit a quadratic inverse despite the
fact that they were not designed with this property in mind. This
observation may not be completely surprising because~\cite{Sun} shows
that good interleavers should require the second degree coefficient to
be relatively large (which works toward satisfying Theorem 3.6) but
bounded by some constraints. This conjecture will be investigated in a future
work.

%
%
\begin{enumerate}[{Example} 1:]
  \item  Let $N = 15120 = 2^4 \cdot 3^3 \cdot 5 \cdot 7 $,
    $f_1 \equiv 11 \pmod{15120}$ and $f_2 \equiv  2 \cdot 3 \cdot 5 \cdot 7 \cdot  m \equiv 210  m \pmod{15120}$,
    where $m$ is any non-negative integer. Let $m=1$.
    Since  $n_{F,2}=1 \geq \max(\lceil \frac{4-2}{2} \rceil,1)$,
    $n_{F,3}=1 \geq \max(\lceil \frac{3-1}{2} \rceil,1)$, 
    $n_{F,5}=1 \geq \lceil \frac{1}{2} \rceil$, $n_{F,7}=1 \geq \lceil \frac{1}{2} \rceil$ and
    $15120$ is even, by Lemma 3.1 and Theorem 3.6, $F(x)$ has two quadratic inverse polynomials. 
    By Algorithms 1 and 2, we can get  $g_{1,1} \equiv 14891 \pmod{15120}$ and $g_{1,2} \equiv 210 \pmod{15120}$, respectively.
    We can also get  $g_{2,1} \equiv g_{1,1} + \frac{15120}{2} \equiv 7331 \pmod{15120} $ 
    and $g_{2,2} \equiv g_{1,2} + \frac{15120}{2} \equiv 7770 \pmod{15120}$ by Algorithm 1.\\
    If $m>1$, there are also two inverses since $m$ only increases $n_{F,p}$, for some $p's$.
    Thus, regardless of the values $m$ and $f_1$, there exist two quadratic inverse polynomials for $F(x)$.
  \item Let $N = 5^3$ and $f_2 \equiv  5  m \pmod{5^3}$, where  $m$ is an integer such that $5 \nmid m$. 
    In this case, regardless of the values $m$ and $f_1$, there are no quadratic inverse polynomial, 
    since $n_{F,5}=1 \not\ge  \lceil \frac{3}{2} \rceil $. 
    However, if $f_2 \equiv 5^2  m \pmod{5^3}$, where $5 \nmid m$, regardless of $m$ and $f_1$,
    there exists one quadratic inverse polynomial since $5^3$ is odd and  
    $n_{{F},5}=2 \ge  \lceil  \frac{3}{2} \rceil $. 
  \item  Let $N = 2^{10}$ and $f_2 \equiv  2^4 \pmod{2^{10}}$. 
    In this case, regardless of the value $f_1$, there exist two inverses since $2^{10}$ is even and
    $n_{F,2} =4 \geq \max(\lceil \frac{10-2}{2} \rceil,1)$. 
    Specifically, if $f_1$ is 1, the two inverses are $G_1(x) = x + 496 x^2 \pmod{2^{10}}$ and $G_2(x) = 513x + 1008 x^2 \pmod{2^{10}}$,
    and if  $f_1$ is 15, the two inverses are $G_1(x) = 751x + 272 x^2 \pmod{2^{10}}$ and $G_2(x) = 239x + 784 x^2 \pmod{2^{10}}$, respectively. 
  %

\end{enumerate}
\section{Conclusion}
  We derived in Theorem 3.6 a necessary and sufficient condition for the
  existence of a quadratic inverse for a quadratic permutation
  polynomial over integer rings. Further, we described a simple
  algorithm (Algorithm 1) to find the coefficients of the quadratic
  inverse polynomial. We also found that almost all good interleavers
  searched in\cite{Sun} admit a quadratic inverse despite the fact that they were
  not designed with this remarkable property in mind. A possible explanation is given. 
  Although this is left for a further investigation, we conjecture that the  
  restriction of quadratic interleavers to admit a quadratic inverse
  does not impair performance when applied to turbo codes. 
  %
  \newpage
\appendix

(A) 
\begin{proof}[Lemma 3.1] 
  \\
  Let $N = \prod\nolimits_{p \in \mathcal{P}}  p^{n_{N,p}} $. $G(x)  = g_1x + g_2 x^2 \pmod{N}$ inverts $F(x)$ at two points: $x = 1$ and $x = 2$
  (in addition, $G(x) $ trivially inverts $F(x)$ at a third point $x = 0$) if and only if the following two congruences  
  have at least one solution set $(g_1, g_2)$.  
  \begin{eqnarray} 
    \label{eq:01}  (G \circ F)(1) =  G(f_1+f_2) & = &     g_1 (f_1 + f_2) + g_2 (f_1 + f_2)^2 \equiv 1 \pmod{N}. \\
    \label{eq:02}  (G \circ F)(2) =  G(2 f_2+ 4 f_2)& =& g_1 ( 2 f_1 + 4 f_2) + g_2 (2 f_1 + 4 f_2)^2 \equiv 2 \pmod{N}.
  \end{eqnarray}
  By multiplying $(2 f_1 + 4 f_2)$ to (\ref{eq:01}) and $(f_1 +f_2)$ to (\ref{eq:02}), we get 
  \begin{eqnarray} 
    \label{eq:03}    g_1 ( 2 f_1 + 4 f_2)(f_1 + f_2) + g_2 ( 2 f_1 + 4 f_2)(f_1 + f_2)^2 & \equiv & 2 f_1 + 4 f_2 \pmod{N}. \\
    \label{eq:04}      g_1 ( 2 f_1 + 4 f_2)(f_1 + f_2) + g_2 (2 f_1 + 4 f_2)^2 (f_1 + f_2) & \equiv  & 2(f_1 + f_2) \pmod{N}.
  \end{eqnarray}
  By subtracting (\ref{eq:03}) from (\ref{eq:04}), 
  \begin{eqnarray} 
    \label{eq:05} 2 g_2 (f_1+f_2)(f_1 + 2 f_2)(f_1+3 f_2)  \equiv -2 f_2 \pmod{N}. 
  \end{eqnarray}
  It can be shown that there exists at least one $g_2$ that satisfies (\ref{eq:05}) as follows.\\
  If either $2 \nmid N$ or $4 | N$, suppose $\gcd(f_1+f_2,N) \neq 1$. Then there is a prime number $p$ such that $ p | (f_1+f_2)$ and $p|N$. 
  However, $p \nmid f_1$ and $p | f_2$ by Corollary 2.7. Thus, $p \nmid (f_1+f_2)$ for all $p$'s such that $p | N$. A contradiction. Therefore $\gcd(f_1+f_2,N) = 1$.
  Similarly, $\gcd(f_1+2 f_2,N) = 1$ and $\gcd(f_1+3 f_2,N) = 1$. Thus, $\gcd((f_1+f_2)(f_1+2f_2)(f_1+3f_2),N) = 1$.
  Consequently, if $2 \nmid N$, $\gcd(2(f_1+f_2)(f_1+2f_2)(f_1+3f_2),N) = 1$ and if $4 | N $,  $\gcd(2(f_1+f_2)(f_1+2f_2)(f_1+3f_2),N) = 2$. \\
  If $2 | N$ and $4 \nmid N$,   $\gcd(2,N) = 2$  
  and  $\gcd((f_1+f_2)(f_1+2f_2)(f_1+3f_2),p) = 1$, where $p \neq 2$ by Corollary 2.7. Thus, $\gcd(2(f_1+f_2)(f_1+2f_2)(f_1+3f_2),N) = 2$. 
  In summary, if $N$ is an even number, we have exactly two solution sets, and if $N$ is an odd number, 
  we have exactly 1 solution set  by Theorem 2.8. \\
  When $N$ is an even number, let $(g_{1,1}, g_{1,2})$ and  $(g_{2,1}, g_{2,2})$ be the solution sets. Then, 
  \begin{eqnarray} 
    \label{eq:06}  g_{1,2}(f_1+f_2)(f_1 + 2 f_2)(f_1+3 f_2)  \equiv -f_2  \pmod{\frac{N}{2}}. 
  \end{eqnarray}
  and $ g_{2,2} \equiv g_{1,2} + \frac{N}{2}  \pmod{N} $ by Theorem 2.8.\\
  When $N$ is an odd number, let $(g_1, g_2) $ be the solution set.
  Then, the congruence (\ref{eq:05}) can be rewritten as \cite{Hardy} 
  \begin{eqnarray} 
   \label{eq:05c}  g_2 (f_1+f_2)(f_1 + 2 f_2)(f_1+3 f_2)  \equiv - f_2 \pmod{N}, 
  \end{eqnarray}
  since $\gcd(2,N) = 1$. \\
  After computing $g_2$ using (\ref{eq:05c}), or $g_{1,2}$, $g_{2,2}$ using (\ref{eq:06}) and Theorem 2.8,  
  we can compute the corresponding $g_1$ or  $g_{1,1}$, $g_{2,1}$ using (\ref{eq:01}) respectively. 
  Specifically, it can be verified that $g_{2,1} \equiv g_{1,1} + \frac{N}{2} \pmod{N} $.
  Thus, for a given quadratic permutation polynomial $F(x)$, we can find at least one quadratic polynomial $G(x)$
  that inverts the polynomial $F(x)$ at three points $x=0, 1, 2$. \\
\end{proof}
(B)
\begin{proof}[Lemma 3.2]
  \begin{enumerate}[{case} a:]
     \item  $2 \nmid N $ \\
        In Lemma 3.1, $F(x)$ is a permutation polynomial. 
        We can thus apply Corollary 2.7 to (\ref{eq:05}) and reducing it to$\pmod{p}$ by Lemma 2.9, where $p$ is a prime number such that $p|N$.
        \begin{eqnarray} 
           2 g_2 \cdot f_1  \cdot f_1 \cdot  f_1  \equiv 0 \pmod{p}. \nonumber
        \end{eqnarray}
        Thus $p|g_{2}$, since $\gcd( 2 \cdot f_1 \cdot f_1 \cdot f_1 ,p) = 1 $, $\forall p$ such that $p | N$.   \\
        By multiplying $(2 f_1 + 4 f_2)^2$ to (\ref{eq:01}) and $(f_1 +f_2)^2$ to (\ref{eq:02}), we get 
        \begin{eqnarray} 
          \label{eq:20}    g_1 ( 2 f_1 + 4 f_2)^2(f_1 + f_2) + g_2 ( 2 f_1 + 4 f_2)^2(f_1 + f_2)^2 & \equiv & (2 f_1 + 4 f_2)^2 \pmod{N}. \\
          \label{eq:21}    g_1 ( 2 f_1 + 4 f_2)(f_1 + f_2)^2 + g_2 (2 f_1 + 4 f_2)^2 (f_1 + f_2)^2 & \equiv  & 2(f_1 + f_2)^2 \pmod{N}.
        \end{eqnarray}
        By subtracting (\ref{eq:21}) from (\ref{eq:20}),
          \begin{eqnarray} 
            \label{eq:22}    g_{1}(f_1+f_2)(2f_1 + 4 f_2)(f_1+3 f_2)  \equiv 2f^2_1 + 12f_1 f_2 + 14 f^2_2 \pmod{N}.
          \end{eqnarray}
        By Lemma 2.9, Corollary 2.7 and  (\ref{eq:22}),      
          \begin{eqnarray} 
            2 g_{1} \cdot f_1  \cdot f_1 \cdot f_1  \equiv 2f^2_1 \pmod{p}. \nonumber  
          \end{eqnarray}
          Thus, if  $p | g_{1}$, then  $p | 2f^2_1 $, which is a contradiction from Corollary 2.7. \\
          By Corollary 2.7, $G(x)$ is a permutation polynomial. 
     \item $ 2 | N  $ and $ 4 \nmid N $   \\ 
        We apply Corollary 2.7. First we prove that $G_1(x)$ and $G_2(x)$ obtained in Lemma 3.1 are permutation polynomials modulo 2.                            
        Since $F(x)$ is a quadratic permutation polynomial, $f_1 + f_2 $ is an odd number from Lemma 2.2. Thus, $(f_1+f_2)^2$ is odd. 
        Let one solution set of Lemma 3.1 be $(g_{1,1},g_{1,2})$.  Suppose $g_{1,1} + g_{1,2} $ is even, i.e.,
        suppose both of $g_{1,1}$ and $g_{1,2}$ are 
        even or odd numbers. Then the LHS of (\ref{eq:01}) becomes an even number.
        A contradiction, since an even number modulo an even number must be an even number but RHS is an odd number and $N$ is even. 
        Therefore, $g_{1,1} + g_{1,2} $ must be an odd number. By Lemma 2.2, $G_1(x)$ is a permutation polynomial modulo 2. 
        Since  $g_{1,1} +  g_{1,2} $ is odd and the second solution set is given as $g_{2,1} \equiv g_{1,1} + \frac{N}{2} \pmod{N}$
        and $g_{2,2} \equiv g_{1,2} + \frac{N}{2} \pmod{N}$, $g_{2,1}+g_{2,2}$ must be odd. 
        Consequently, $G_2(x)$ is a permutation polynomial modulo 2.  
        For $p$'s such that $p \neq 2$, using a similar argument as in case (a), 
        it can be easily verified that $G_1(x)$ and $G_2(x)$ are  permutation polynomials. 
     \item  $4|N$  \\
        We apply Corollary 2.7. First we prove that $G_1(x)$ and $G_2(x)$ obtained in Lemma 3.1
        are permutation polynomials modulo $2^n$ where $n \geq 2$. \\  
        $f_2$ is even by Corollary 2.5 and $\frac{N}{2}$ is even since $4|N$. 
        By Corollary 2.5, $f_1+f_2$, $f_1+2f_2$, $f_1+3f_2$ are all odd numbers. Thus  $(f_1+f_2)(f_1 + 2 f_2)(f_1+3 f_2)$ is an odd number.
        Consequently, $g_{1,2}$ must be an even number  in (\ref{eq:06}) by reducing it $\pmod{2}$ and using Lemma 2.9.
        From (\ref{eq:01}), since $f_1+f_2$ is odd and $g_{1,2}(f_1+f_2)^2$ is even, $g_{1,1}$ must be an odd number.
        Finally, $g_{2,1} \equiv g_{1,1} + \frac{N}{2} \pmod{N}$ is an odd number since $g_{1,1}$ is an odd number, and 
        $g_{2,2} \equiv g_{1,2} + \frac{N}{2} \pmod{N}$ is an even number since $g_{1,2}$ is an even number.
        Consequently, by Corollary 2.5, $G_1(x)$ and $G_2(x)$ are permutation polynomials modulo $2^n$, where $n \geq  2$.
        For $p$'s such that $p \neq 2$, using a similar argument as in case (a), 
        it can be easily verified that $G_1(x)$ and $G_2(x)$ are permutation polynomials. 
    \end{enumerate}
\end{proof}
(C)
\begin{proof}[Lemma 3.3]
  \begin{enumerate}[{case} a:]
     \item $2 \nmid N$ \\
       Suppose that $n_{{F,p}} < n_{{N,p}}$ and $n_{{F,p}} < n_{{G,p}}$, where $p$ is a prime number such that $p|N$.
       From Lemma 2.9 and (\ref{eq:05c})
       \begin{eqnarray}  
          0  \equiv -f_2 \pmod{ p^{ \min{( n_{G,p}, n_{N,p})} } }.  \nonumber
       \end{eqnarray}
       A contradiction. \\
       Now suppose that $n_{{F,p}} < n_{{N,p}}$ and $n_{{G,p}} < n_{{F,p}}$, again, from Lemma 2.9 and (\ref{eq:05c})
       \begin{eqnarray} 
          g_2 \cdot f_1 \cdot f_1 \cdot f_1  \equiv 0 \pmod{p^{n_{F,p}}}. \nonumber
       \end{eqnarray}
       The LHS cannot be $0$, since  $\gcd( f_1 \cdot f_1 \cdot f_1 , p^{n_{F,p}} )=1$ by Corollary 2.7.
       A contradiction. Thus $n_{{G,p}} = n_{{F,p}}$.\\
       If $n_{{F,p}} \geq n_{{N,p}}$, from Lemma 2.9 and (\ref{eq:05c}), 
       \begin{eqnarray} 
         g_2 \cdot f_1 \cdot f_1 \cdot f_1  \equiv 0 \pmod{p^{n_{N,p}}}, \nonumber
       \end{eqnarray}
       which forces $n_{{G,p}} \geq n_{{N,p}}$. 
    \item $ 2 | N $ and $ 4 \nmid N $ \\
       Using a similar argument as above, it is easily verified by using Lemma 2.9 and (\ref{eq:06}).
    \item $ 4 | N $ \\
       Using a similar argument as above, it is easily verified by using Lemma 2.9 and (\ref{eq:06}).
  \end{enumerate}
\end{proof}
(D)
\begin{proof}[Lemma 3.4]
    \\   ( $\Longrightarrow$ ) \\
    Define $T_0 (x) = T(x) = t_1 x + t_2 x^2 + t_3 x ^3 + t_4 x^4 \pmod{N}$ 
    and $ T_n (x) = T_{n-1}(x+1) - T_{n-1}(x)$, $\forall n \geq 1$.
    If $T(x) \equiv 0 \pmod{N}$, $\forall x \in [0,N-1]$ then 
    $T_n (x) \equiv 0 \pmod{N}$, $\forall x \in [0,N-1]$, $\forall n \geq 0$.
    After some computation, it can be easily shown that 
    \begin{eqnarray} 
       T_1 (x) & = &  (t_1 + t_2 + t_3 + t_4)  + (2 t_2 + 3 t_3 + 4 t_4)x + (3 t_3 + 6 t_4)x^2 +  4 t_4 x^3  \equiv 0  \pmod{N}.   \nonumber \\
       T_2 (x) & = &  ( 2 t_2 + 6 t_3 + 14 t_4) + (6 t_3 + 24 t_4 )x + 12 t_4 x^2   \equiv 0  \pmod{N}.     \nonumber \\
       T_3 (x) & = &  (  6 t_3 + 36 t_4 ) + 24 t_4 x   \equiv 0  \pmod{N}.   \nonumber
    \end{eqnarray}
    Consequently, in order to ensure $T_3 (x) \equiv 0 \pmod{N}$ for $x \in [0,N-1] $,   
    \begin{eqnarray} 
       24 t_4 \equiv 0 \pmod{N}.     \nonumber  \\
       6 t_3 + 36 t_4  \equiv 0 \pmod{N}.     \nonumber
    \end{eqnarray}
   ( $\Longleftarrow$ ) \\
    Define $T_0(x), T_1(x), T_2(x) and T_3(x)$ as above. 
    Then by assumption, $T_3(x) \equiv 0 \pmod{N}$, $\forall x \in [0,N-1]$, and $T_2(0) = T(2) -2 T(1) +  T(0) \equiv 0 \pmod{N} $.  
    By induction, $T_2(x) \equiv 0 \pmod{N}$, $\forall x \in [0,N-1]$  since $T_2(x+1) = T_2(x) + T_3(x) $.  
    By the same procedure, $T_1(x) \equiv 0 \pmod{N}$ and $T(x) =T_0(x) \equiv 0 \pmod{N}$.  
\end{proof}
(E) 
\begin{proof}[Theorem 3.5]
   \\
   $(G \circ F)(x) \equiv x \pmod{N}$  if and only if $G(x)$ is the quadratic inverse polynomial of $F(x)$. 
   \begin{eqnarray} 
      (G \circ F)(x) & \equiv & g_1 (f_1x+f_2x^2) + g_2 (f_1x+f_2x^2)^2  \pmod{N}    \nonumber \\
                     & \equiv & f_1 g_1 x + (f_2 g_1+ {f_{1}}^2 g_2)x^2 + 2 f_1 f_2 g_2 x^3 + {f_2}^2 g_2 x^4 \pmod{N}    \nonumber \\
                     & \equiv & x \pmod{N}. \nonumber
   \end{eqnarray}
   Thus,  $G(x)$ is the quadratic inverse polynomial of $F(x)$ if and only if the following condition is satisfied. 
   \begin{eqnarray} 
     \label{eq:30}   (f_1 g_1-1) x + (f_2 g_1 + {f_1}^2 g_2)x^2 + 2 f_1 f_2 g_2 x^3 + {f_2}^2 g_2 x^4 \equiv 0 \pmod{N}  
   \end{eqnarray}
   Let $ T(x) = (G \circ F)(x)-x = (f_1 g_1-1) x + (f_2 g_1+ {f_1}^2 g_2)x^2 + 2 f_1 f_2 g_2 x^3 + {f_2}^2 g_2 x^4 $. 
   By Lemma 3.1, $T(0) = G(F(0))-0 \equiv 0 \pmod{N} $, $ T(1) = G(F(1))-1 \equiv 0 \pmod{N} $, $T(2) =  G(F(2))-2 \equiv 0 \pmod{N}$.
   Applying Lemma 3.4, we get
   \begin{eqnarray} 
     24 {f_2}^2 g_2 & \equiv & 0  \pmod{N}.    \nonumber \\
     36 f^2_2 g_2 + 12 f_1 f_2 g_2  & \equiv & 0 \pmod{N}.    \nonumber     
   \end{eqnarray}
   These can be further reduced to 
   \begin{eqnarray} 
     12 f_2 g_2 \equiv  0 \pmod{N},     \nonumber     
   \end{eqnarray} 
   since $\gcd(f_1 +f_2, N) = 1$, by Corollary 2.7. 
\end{proof}
(F)
\begin{proof}[Theorem 3.6]
   \\
   ( $\Longrightarrow$ ) \\
   By Lemmas 3.1 and 3.2, a quadratic permutation polynomial $F(x)$ has 
   at least one quadratic permutation polynomial $G(x) = g_1 x + g_2 x^2
   \pmod{N}$ that inverts  $F(x)$ at three points
   $x=0, 1, 2$. Since it is required for a quadratic inverse polynomial to
   invert $F(x)$ at these points, we only need to check
   whether $G(x)$ is a quadratic inverse polynomial or not.\\
   We show that if $G(x)$ is a quadratic inverse polynomial, then the condition on $n_{F,p}$, where $p=2$, holds.
   The conditions for $n_{F,p}$, where $p \neq 2$, can be done similarly.  \\
   If $n_{N,2} = 0,1$, whether $G(x)$ is a quadratic inverse or not, $n_{F,2} \geq 0$ trivially holds and
   this is why we do not need to determine $n_{G,2}$ in Lemma 3.3, case (b) when $n_{N,2} = 1$.\\ 
   If $n_{N,2} = 2,3,4$, by Corollary 2.7, $n_{F,2} \geq 1$, since $F(x)$ is a permutation polynomial. \\
   Suppose that  $G(x)$ is a quadratic inverse polynomial but $n_{F,2} < \left\lceil \frac{n_{N,2}-2}{2} \right\rceil $, for $n_{N,2} \geq 5$. 
   Since $G(x)$ is a quadratic inverse polynomial,  $12 f_2 g_2 \equiv 0 \pmod{N}$ holds by Theorem 3.5, i.e., 
   $\prod\nolimits_{p \in \mathcal{P}}  p^{n_{N,p}} | (2^2 \cdot 3 \cdot \prod\nolimits_{p \in \mathcal{P}}  p^{n_{F,p}}  \cdot  \prod\nolimits_{p \in \mathcal{P}}  p^{n_{G,p}})$. \\
   We divide it into two cases. 
   \begin{enumerate}
      \item $n_{N,2}$ is odd \\
        $\left\lceil \frac{n_{N,2}-2}{2} \right\rceil  =   \frac{n_{N,2}-1}{2} $, thus $n_{F,2}  \leq \frac{n_{N,2}-1}{2}-1$. By Lemma 3.3, 
        $n_{G,2} = n_{F,2}$, thus $2 + n_{F,2} + n_{G,2} \leq n_{N,2}-1 < n_{N,2}  $, which is a contradiction since
        $N| 12 f_2 g_2 $ implies $n_{N,2} \leq 2+ n_{F,2} + n_{G,2} $.    
      \item $n_{N,2}$ is even \\
        $\left\lceil \frac{n_{N,2}-2}{2} \right\rceil  =   \frac{n_{N,2}-2}{2} $, thus $n_{F,2}  \leq \frac{n_{N,2}-2}{2}-1$. By Lemma 3.3, 
        $n_{G,2} = n_{F,2}$, thus $2 + n_{F,2} + n_{G,2} \leq n_{N,2}-2 < n_{N,2}  $, which is a contradiction since
        $N| 12 f_2 g_2 $ implies $n_{N,2} \leq 2+ n_{F,2} + n_{G,2} $. 
   \end{enumerate}
   Similarly, it can be shown that if $G(x)$ is a quadratic inverse polynomial, then the conditions on $n_{F,p}$, where $p \neq 2$ is satisfied. 
   %
   %
   ( $\Longleftarrow$ ) \\
   We show that if the conditions on $n_{F,p}$ holds, then $12 f_2 g_2 \equiv 0 \pmod{N}$, i.e.,
   $\prod\nolimits_{p \in \mathcal{P}}  p^{n_{N,p}} | (2^2 \cdot 3 \cdot     \prod\nolimits_{p \in \mathcal{P}}  p^{n_{F,p}}  \cdot  \prod\nolimits_{p \in \mathcal{P}}  p^{n_{G,p}})$ holds. 
   We only show that if the condition on $n_{F,p}$, where $p=2$, is satisfied,  $ 2^{n_{N,2}} | (2^2 \cdot  2^{n_{F,2}} \cdot 2^{n_{G,2}}) $ holds 
   and the case where $p \neq 2$ can be done similarly. \\
   We divide it into three cases. 
   \begin{enumerate}
      \item $n_{N,2}=0,1$ \\
         If $n_{F,2} \geq 0$, then $n_{N,2} \leq 2+n_{F,2}$  holds. Thus  $ 2^{n_{N,2}} | (2^2 \cdot 2^{n_{F,2}} \cdot 2^{n_{G,2}} ) $   
      \item $n_{N,2}=2,3,4$ \\
         If  $n_{F,2} \geq 1$, then as required by Lemma 3.3,  $n_{G,2} \geq 1$. Thus, $n_{N,2} \leq 2+ n_{F,2} + n_{G,2} $ holds and consequently
         $ 2^{n_{N,2}} | (2^2 \cdot  2^{n_{F,2}} \cdot 2^{n_{G,2}})  $   
      \item $n_{N,2} \geq 5$ \\
         If $n_{N,2} \geq 5$, then $\left\lceil \frac{n_{N,2}-2}{2} \right\rceil > 1 $.
         By Lemma 3.3, if $n_{N,2}-1> n_{F,2} \geq \left\lceil \frac{n_{N,2}-2}{2} \right\rceil$, then
         $n_{G,2} = n_{F,2} $. Consequently,  if $n_{N,2}$ is even,  $ 2 + n_{F,2} + n_{G,2} =  2 + 2 \cdot n_{F,2} \geq 2 + n_{N,2}-2 = n_{N,2}$,
         and if $n_{N,2}$ is odd, $ 2 + n_{F,2} + n_{G,2} =  2 + 2 \cdot n_{F,2}  \geq 2 + n_{N,2}-1  > n_{N,2}$.
         Thus $ 2^{n_{N,2}} | (2^2 \cdot  2^{n_{F,2}} \cdot 2^{n_{G,2}} ) $.  If $n_{F,2} \geq n_{N,2} -1 $, by Lemma 3.3, 
         $n_{G,2} \geq n_{N,2} -1 $. Thus $2 + n_{F,2} + n_{G,2} \geq  2 \cdot n_{N,2} > n_{N,2}$
         and consequently $ 2^{n_{N,2}} | (2^2 \cdot  2^{n_{F,2}} \cdot 2^{n_{G,2}} ) $.
   \end{enumerate}
\end{proof}
%
%
%
%
%

%



\end{document}